\documentclass[12pt,showpacs]{revtex4}
\usepackage{amssymb,epsf}
\usepackage{latexsym}
\usepackage{color}
\begin{document}
\title{Quantum corrections for a black hole in an asymptotically safe gravity
with higher derivatives}

\author{Mubasher Jamil}
\email{mjamil@camp.nust.edu.pk} \affiliation{Center for Advanced
Mathematics and Physics, National University of Sciences and
Technology, H-12, Islamabad, Pakistan}

\author{Farhad Darabi}
\email{f.darabi@azaruniv.edu, } \affiliation{Department of Physics,
Azarbaijan University of Tarbiat Moallem, Tabriz, Iran \\
Research Institute for Astronomy and Astrophysics of Maragha (RIAAM), Maragha 55134-441, Iran\\
Corresponding Author}

\date{\today}
\begin{abstract}
\textbf{Abstract:} 
By using the quantum tunneling approach over semiclassical approximations, we study the quantum corrections to the Hawking temperature, entropy and Bekenstein-Hawking entropy-area relation for a black hole in an asymptotically safe gravity with higher derivatives. The leading and non leading corrections
to the area law are obtained.
\end{abstract}
\maketitle
\newpage
\section{Introduction}
The quantum corrections to the black hole solutions are of current interest
and have been calculated by different methods\cite{Schw, Kerr, BTZ}. String theory, loop quantum gravity and noncommutative geometry show that in the entropy-area relation the leading order correction should be of log-area type \cite{Str1, Str2, Str3, Str4, Str5, Med, Set1,
Set2, Set3, Alex1, Alex2, Modak}. On the other hand, generalized uncertainty principle (GUP) and modified dispersion relations (MDRs) provide perturbational framework for such modifications \cite{Noz, Sef, Far, Set4, Set5}. 
In the present paper, we use quantum tunneling approach over semiclassical approximations to calculate the quantum corrections to the Hawking temperature and entropy of black hole solutions. In a recent work \cite{Jamil}, we have calculated such quantum corrections for a Braneworld black hole corresponding to the solution presented in \cite{roy}. Here, we calculate the quantum corrections to the Hawking temperature and entropy of black hole solutions in an asymptotically safe gravity theory with higher derivatives \cite{cai}.

\section{Hawking entropy-area relation}

To calculate the corrections to the entropy and temperature of a
black hole, we use the Hamilton-Jacobi method to compute the
imaginary part of the action outside the semi-classical
approximation admitting all possible quantum corrections . The
expression of quantum correction of a general function $S(r,t)$
expanded in the series in powers of $\hbar$ is \cite{jhep}
\begin{equation}\label{3}
S(r,t)=S_0(r,t)+\hbar S_1(r,t)+\hbar^2S_2(r,t)+\ldots=S_0(r,t)
+\sum_i\hbar^iS_i(r,t).
\end{equation}
Here $S_0$ is the semiclassical entropy while the remaining terms
are the higher order corrections to $S_0$. On the considerations of
dimensional analysis, the above expression (\ref{3}) takes the form
\cite{jhep}:
\begin{equation}\label{4}
S(r,t)=S_0(r,t)+\sum_i\frac{\hbar^i}{M^{2i}}S_0(r,t)=S_0(r,t)\Big(
1+\sum_i\alpha_i \frac{\hbar^i}{M^{2i}}\Big),
\end{equation}
where $M$ is the mass of the black hole and $\alpha$'s are the
correction coefficients.
The modified form of the temperature of the BH can be written as
\cite{jhep}
\begin{equation}\label{5a}
T=T_H\Big( 1+\sum_i\alpha_i \frac{\hbar^i}{M^{2i}} \Big)^{-1},
\end{equation}
where $T_H$ is the standard semiclassical Hawking temperature and
the terms with $\alpha_i$ are corrections due to quantum effects. If
we consider $\alpha_i$ in terms of a singular dimensionless
parameter $\beta$ such that $\alpha_i=\beta^i$ then we have
\begin{equation}\label{5aa}
1+\sum_i\alpha_i \frac{\hbar^i}{M^{2i}}=\Big(
1-\frac{\beta\hbar}{M^2} \Big)^{-1}.
\end{equation}
On comparing (\ref{5aa}) with (\ref{5a}), we obtain
\begin{equation}\label{5b}
T=T_H\Big( 1-\frac{\beta\hbar}{M^2} \Big).
\end{equation}
Motivated by a recent work \cite{cai}, we use the
static spherically symmetric spacetime in an asymptotically safe
high-derivative gravity
\begin{equation}\label{1a}
ds^2=-f(r)dt^2+\frac{dr^2}{f(r)}+r^2d\Omega^2,
\end{equation}
where
\begin{equation}\label{1}
f(r)=1-\frac{2M}{r}G_N+\frac{2M}{r^3}G_N^2\tilde{\xi}.
\end{equation}
The black hole temperature is
\begin{eqnarray}\label{2}
T_H=\hbar\frac{f'(r)}{4\pi}|_{r=r_e}&=&\hbar
\frac{2MG_N}{r_e^2}\Big(1-\frac{3G_N\tilde\xi}{r_e^2}\Big)
\\&=&\hbar \frac{1}{8\pi G_N M}\Big( 1-\frac{3\tilde{\xi}}{4G_N M^2}
\Big).
\end{eqnarray}
Here $r_e=2G_N M$ is the event horizon of the black hole.

\begin{equation}\label{3a}
T=T_H\Big( 1-\frac{\beta\hbar}{M^2} \Big)
\end{equation}
Making use of the first law of thermodynamics, the semiclassical
entropy is given by
\begin{equation}\label{5}
S_0=\int
\frac{1}{T_H}dM=\frac{8\pi}{\hbar}\int\frac{r_e^2}{2r_e-3\tilde\xi}dM.
\end{equation}
Using the differential of mass in (\ref{5}), we get
\begin{equation}\label{5c}
S_0=\frac{4\pi}{\hbar G_N}\int\frac{r_e^2}{2r_e-3\tilde\xi}dr_e.
\end{equation}
The above integral gives
\begin{eqnarray}\label{5d}
S_0&=&\frac{\pi}{G_N\hbar}r_e^2+\frac{3}{G_N\hbar}r_e\tilde\xi
+\frac{9}{2\hbar}\pi\tilde\xi^2\ln(2r_e-3\tilde\xi)\nonumber\\
&\simeq&\frac{\pi}{G_N\hbar}r_e^2+\frac{3}{G_N\hbar}r_e\tilde\xi
+O(\tilde\xi^2).
\end{eqnarray}
Assuming $A$ is the area of the event horizon,
\begin{equation}\label{10a}
A=\int\sqrt{g_{22}g_{33}}dx^2dx^3=4\pi r_e^2.
\end{equation}
Thus (\ref{5d}) can be written as
\begin{equation}\label{11}
S_0\simeq\frac{A}{4\hbar G_N}+\frac{3}{G_N}\pi\tilde\xi r_e.
\end{equation}
Thus we obtain the Hawking entropy-area relation for the black hole
under consideration.

\subsection{Hawking Temperature Corrections}

In this section, we find the correction to the Hawking temperature
as a result of quantum effects for the black hole. The expression
for the semiclassical Hawking temperature, (\ref{2}) turns out to be
\begin{eqnarray}\label{2a}
T_H&=& \frac{\hbar}{r_e}\Big(1-\frac{3G_N\tilde\xi}{r_e^2}\Big).
\end{eqnarray}
The corrected temperature, from (\ref{5b}) turns out
\begin{equation}\label{5q}
T=T_H\Big( 1-\frac{\beta\hbar r_e^2}{4G_N^2} \Big).
\end{equation}
Using (\ref{2a}) and (\ref{5q}), we obtain the quantum correction
for the temperature
\begin{equation}\label{5e}
T=\frac{\hbar}{r_e}\Big(1-\frac{3G_N\tilde\xi}{r_e^2}\Big)\Big(
1-\frac{\beta\hbar r_e^2}{4G_N^2} \Big).
\end{equation}

\subsection{Entropy Corrections}

Here we calculate the quantum corrections to the entropy of the
black hole. In terms of the horizon radius, the corrected form of
the entropy (\ref{4}) is given by
\begin{equation}\label{4a}
S=S_0\Big( 1+\sum_i
\frac{\alpha_i\hbar^i(2G_N)^{2i}}{r_e^{2i}}\Big),
\end{equation}
and similarly the corrected form of the Hawking temperature can be
written as
\begin{equation}\label{4c}
T=T_H\Big( 1+\sum_i
\frac{\alpha_i\hbar^i(2G_N)^{2i}}{r_e^{2i}}\Big)^{-1}.
\end{equation}
Again using the first law of thermodynamics,
\begin{equation}\label{4d}
S=\int\frac{1}{T_H}\Big( 1+\sum_i
\frac{\alpha_i\hbar^i(2G_N)^{2i}}{r_e^{2i}}\Big)dM,
\end{equation}
which can be written in expanded form as
\begin{eqnarray}\label{4e}
S&=&\int\frac{1}{T_H}dM+\int\frac{\alpha_1\hbar(2G_N)^{2}}{T_Hr_e^{2}}dM
+\int\frac{\alpha_2\hbar^2(2G_N)^{4}}{T_Hr_e^{4}}dM+\ldots,\nonumber\\
&=&I_1+I_2+I_3+\ldots,
\end{eqnarray}
where the first integral $I_1$ has been evaluated in (\ref{11}) and
$I_2$, $I_3,\ldots$ are quantum corrections. Thus,
\begin{equation}\label{4f}
I_2=2^4\pi\alpha_1\hbar G_N\int\frac{dr_e}{2r_e-3\tilde\xi},
\end{equation}
and
\begin{equation}\label{4g}
I_3=2^6\pi\alpha_2\hbar
^2G_N^3\int\frac{dr_e}{r^2_e(2r_e-3\tilde\xi)}.
\end{equation}
In general, we can write
\begin{equation}\label{4h}
I_k=2^{2k}\pi\alpha_{k-1}\hbar^{k-1}
G_N^{2k-3}\int\frac{dr_e}{r^{2k-4}_e(2r_e-3\tilde\xi)}.
\end{equation}
Therefore, the entropy with quantum corrections is given by
\begin{eqnarray}\label{4j}
S&=& \frac{4\pi}{\hbar G_N}\int\frac{r_e^2}{2r_e-3\tilde\xi}dr_e+
2^4\pi\alpha_1\hbar G_N\int\frac{dr_e}{2r_e-3\tilde\xi}+\sum_{k>2}
2^{2k}\pi\alpha_{k-1}\hbar^{k-1}
G_N^{2k-3}\int\frac{dr_e}{r^{2k-4}_e(2r_e-3\tilde\xi)}.\nonumber\\
\end{eqnarray}
After the integrals are evaluated, (\ref{4j}) takes the form
\begin{eqnarray}\label{4k}
S&\simeq&\frac{\pi}{G_N\hbar}r_e^2+\frac{3}{G_N\hbar}r_e\tilde\xi
+2^3\alpha_1\hbar
G_N\ln(2r_e-3\tilde\xi)+\frac{2^6\alpha_2\hbar^2\pi G_N^3}{3^2}\Big[
\frac{3}{r_e\tilde\xi}+\frac{2}{\tilde\xi^2}\ln(2-\frac{3\tilde\xi}{r_e})
\Big]\nonumber\\&&+\sum_{k>3}\frac{2^{2k}\pi\alpha_{k-1}\hbar^{k-1}
G_N^{2k-3}r_e^{5-2k}}{\tilde\xi(6k-15)}{}_{2}F_{1}
\Big(5-2k,1,6-2k,\frac{2r_e}{3\tilde\xi}\Big).
\end{eqnarray}
The expression can be written in terms of the area $A$, using (\ref{10a}), as follows
\begin{eqnarray}\label{4l}
S&\simeq&\frac{A}{4G_N\hbar}+\frac{3}{G_N\hbar}\sqrt{\frac{A}{4\pi}}\tilde\xi
+2^3\alpha_1\hbar
G_N\ln(2\sqrt{\frac{A}{4\pi}}-3\tilde\xi)+\frac{2^6\alpha_2\hbar^2\pi G_N^3}{3^2}\Big[
\frac{3\sqrt{4\pi}}{\sqrt{A}\tilde\xi}+\frac{2}{\tilde\xi^2}\ln(2-\frac{3\sqrt{4\pi}\tilde\xi}{\sqrt{{A}}})
\Big]\nonumber\\&&+\sum_{k>3}\frac{2^{2k}\pi\alpha_{k-1}\hbar^{k-1}
G_N^{2k-3}\left(\frac{A}{4\pi}\right)^\frac{5-2k}{2}}{\tilde\xi(6k-15)}{}_{2}F_{1}
\Big(5-2k,1,6-2k,\frac{2\sqrt{A}}{3\sqrt{{4\pi}}\tilde\xi}\Big),
\end{eqnarray}
where ${}_2F_{1}$ is the hypergeometric function. Clearly the first
two terms in the above expansion is the usual Hawking entropy-area relation
(\ref{11}). The third term is the leading term which is logarithmic as has been found earlier for other geometries using different methods \cite{Str1, Str2, Str3, Str4, Str5, Med, Set1,
Set2, Set3, Alex1, Alex2, Modak}. The higher order corrections are 
combinations of hypergeometric functions.
\newpage
\section{Conclusion}

There are several approaches such as string theory,
loop quantum gravity, noncommutative geometry, modified dispersion relations and generalized uncertainty principle to find quantum gravitational corrections of black hole entropy-area relation. In this paper, using the quantum tunneling approach over semiclassical approximations, we have studied the quantum corrections to the thermodynamical quantities like Hawking temperature, entropy and Bekenstein-Hawking
entropy-area relation of a black hole in an asymptotically safe gravity
with higher derivatives. The first two terms are the semiclassical entropy-area relation, while the third term is the leading logarithmic term.  The higher order corrections involve the inverse of the square root of area term and
are combinations of hypergeometric functions.

\section*{Acknowledgment}
This work has been supported by ``Research Institute for
Astronomy and Astrophysics of Maragha (RIAAM)'', Iran.

\end{document}